\documentclass[journal]{IEEEtran}

\usepackage{graphicx,cite}
\usepackage{amsmath, mathtools}
\usepackage{url}
\usepackage{caption}
\usepackage{subcaption}
\usepackage{amssymb}
\usepackage{array}
\usepackage{fixltx2e}
\usepackage{amsfonts}
\usepackage{lineno}
\usepackage{dsfont}
\usepackage{bbm}
\usepackage{multirow}
\usepackage{booktabs}
\usepackage{bigdelim}
\usepackage{soul}
\usepackage[dvips]{color}
\usepackage{epsf}
\usepackage{times}
\usepackage{epsfig}
\usepackage{amsxtra}
\usepackage{algorithm}
\usepackage[normalem]{ulem}
\usepackage{algpseudocode}

\begin{document}

\title{Co-Primary Multi-Operator Resource Sharing for Small Cell Networks}

\author{\IEEEauthorblockN{Petri Luoto, Pekka Pirinen, Mehdi Bennis, Sumudu Samarakoon, Simon Scott and Matti Latva-aho} \\ 
\IEEEauthorblockA{Centre for Wireless Communications
University of Oulu, Finland, \\ email: \url{{petri.luoto,pekka.pirinen,bennis,sumudu,simon.scott,matti.latva-aho}@ee.oulu.fi}}
\thanks{This research was supported by the Finnish Funding Agency for Technology and Innovation (TEKES), Nokia Networks, Anite Telecoms, Huawei Technologies, Broadcom Communications Finland, Elektrobit Wireless Communications and Infotech Oulu Graduate School. Kari Horneman, Ling Yu and Eric Galloix from Nokia Networks earn special thanks for
suggesting this research direction and for giving invaluable feedback.}
}

\IEEEtitleabstractindextext{%
\begin{abstract}
To tackle the challenge of providing higher data rates within limited spectral resources we consider the case of multiple operators sharing a common pool of radio resources. Four algorithms are proposed to address co-primary multi-operator radio resource sharing under heterogeneous traffic in both centralized and distributed scenarios. The performance of these algorithms is assessed through extensive system-level simulations for two indoor small cell layouts. It is assumed that the spectral allocations of the small cells are orthogonal to the macro network layer and thus, only the small cell traffic is modeled. The main performance metrics are user throughput and the relative amount of shared spectral resources. The numerical results demonstrate the importance of coordination among co-primary operators for an optimal resource sharing. Also, maximizing the spectrum sharing percentage generally improves the achievable throughput gains over non-sharing.
\end{abstract}

}

\maketitle

\IEEEdisplaynontitleabstractindextext

\IEEEpeerreviewmaketitle

\section{Introduction}
\IEEEPARstart{W}{ireless} cellular systems are experiencing a growth data rate demands from users and it is expected that this trend will continue to speed up in the near future. Even though the fourth generation (4G) is still in its infancy, yet growing rapidly, the interest has already moved toward fifth generation (5G) networks. The continuing growth in demand for better coverage and capacity enhancements is pushing the industry to look ahead at how networks can meet future extreme capacity and performance demands \cite{ericsson13,ericsson14}.

5G mobile communication systems are expected to revolutionize everything seen so far in wireless systems. The requirements for 5G vary by application but will include data rates ranging from very low sensor data to very high video content delivery, stringent low latency requirements, low energy consumption, and high reliability \cite{5Gvision}. All of these technological requirements are expected to be achieved while keeping similar or lower cost than today's technologies. 5G is likely to integrate enhancements in legacy radio access technologies with new developments in the areas of multiple access, waveform design, interference management, access protocols, network architecture and virtualization, massive MIMO, full-duplex radio technology, low latency, device-to-device (D2D) and machine type communication (MTC), etc \cite{metis}.

In addition to radio access technology advances, network capacity and connectivity can be improved by network densification (mainly via small cell deployment) and by harnessing broader spectrum allocations. In addition to small cell deployments, there are many other techniques and systems that can improve coverage and data rates, in densely populated indoor environments. These techniques include the deployment of radio remote heads (RRHs), distributed antenna systems (DAS), WiFi access points, etc. The use of LTE small cells offers several advantages over such systems. Compared to DAS, LTE small cells are both cheaper and less complex to deploy \cite{Zhang_femtocells}, and compared to WiFi, LTE small cells offer better performance, more efficient use of resources, and are well designed to support a substantial number of users \cite{huawei1}.

Future networks are expected to include innovative ways of sharing both content and spectrum. This can be seen from observing current trends \cite{nsnCoPSS}. Mobile network operators (MNOs), which we will refer to as operators (OPs) hereafter in current wireless communication networks have commonly acquired exclusive usage rights for certain frequency bands and have little incentives to share it with other operators, despite significant research and regulatory efforts. This might be due to the lack of joint technological and business consideration. However, due to high cost and spectrum scarcity it can be expected that efficient use of spectrum in 5G networks will rather rely on sharing than exclusive licenses.

A recently proposed novel spectrum sharing mechanism towards 5G systems is the so-called co-primary spectrum sharing (CoPSS), where any OP is allowed to utilize shared spectrum allocated for 5G cellular systems. In \cite{metis2}, CoPSS is defined as a spectrum access model where primary license holders agree on the joint use of (or parts of) their licensed spectrum. This would be possible in the small cells domain only where base stations coverage is similar to today's WiFi access points and the frequency band is dedicated to small cell use. Depending on the expected time-frame for 5G roll-out, there are different views on the 5G system concept. The next World Radiocommunication Conference 2015 (WRC-15) will be quite important in setting the directions towards the next standard because it is evident that the next generation standard must be open enough to allow new spectrum sharing methods and drastically new technologies not even known during the development phase \cite{wrc}. Lot of discussion is going on regarding spectrum sharing at 3.5 GHz band in small cells which is in agenda of WRC-15 \cite{nsnCoPSS}.

In \cite{CoPPS_business}, enabling/limiting factors for CoPSS are discussed. Therein, the current scarcity of spectrum and new business potential, especially in hotspots and small cells, are seen as enabling factors for CoPSS. The limited availability of suitable spectrum for sharing, a low level of technical/business knowledge among OPs regarding CoPSS, and a lack of rules to coordinate sharing between OPs with similar customer profiles, are seen as limiting factors. The findings suggest that substantial further research is required, not only from a technical perspective, but also from business perspective.

Multi-operator spectrum sharing has been considered in many research papers over the years \cite{ss0c,ss1c,ss2c,ss2j}. In \cite{saphyre}, various aspects of inter-operator resource sharing have been studied such as analyzing and developing new self-organizing physical layer resource sharing models, analyzing efficient co-ordination mechanisms and developing a framework for infrastructure sharing. In \cite{saphyre2}, the potential gain of spectrum sharing between cellular operators in terms of network efficiency is investigated. In \cite{ss1j}, inter-operator sharing of cellular resources including capacity, spectrum and base stations is investigated. Therein, realistic sharing processes and architecture are proposed compatible with LTE.

To the best of our knowledge, the potential of CoPSS in LTE indoor multi-operator small cell base station (SBS) network has not been investigated. In this paper, CoPSS at the physical resource block (PRB) level is studied. Spectrum sharing at PRB level is challenging because different OPs' SBSs have to be synchronized. However this type of spectrum sharing guarantees more efficient utilization of the spectrum. Coarser granularity component carrier level resource sharing may be a more practical approach when multi-operator networks are not jointly synchronized.

The main focus of this work is on CoPSS between SBSs belonging to different OPs. A dense indoor network deployment, consisting of multiple SBSs per building operated by three independent OPs, is considered. Traffic in the network is heterogeneous, i.e. a mix of full buffer and continuous constant rate traffic. Four CoPSS algorithms are proposed, and their performance is evaluated. These algorithms enable CoPSS when SBSs are not using 100\% of their bandwidth. A given SBS is not fully utilizing its bandwidth when it can provide the required data rate for all the users without using 100\% of its bandwidth, i.e. minimum bandwidth usage by SBS is ensured by utilizing the maximum transmission power and the highest order modulation and coding scheme (MCS) possible for all transmissions. From an energy efficiency perspective it may be beneficial for a SBS to utilize its full bandwidth for all transmissions. However, from CoPSS point of view it is more beneficial to keep the bandwidth usage to a minimum. Typically SBSs are placed in densely populated environments without frequency planning, and have time-variant traffic profiles. The aim is to use spectrum more efficiently, in order to reduce future spectrum requirements and increase the capacity of small cell networks.

Each SBS OP has its own dedicated spectrum, and each OP can define a percentage of how much spectrum they are willing to share. The idea of CoPSS is that spectrum is shared orthogonally and equally between operators. This way interference can be avoided and spectrum utilization is maximized. In three proposed algorithms, unused resources are shared equally between overloaded OPs for a given time instant, short term fairness among overloaded SBSs can be guaranteed. However, long term fairness between OPs cannot be guaranteed, i.e. the equal amount of the average loaned/rented spectrum usage over a given time period. Therefore, there is a need for a  spectrum sharing framework that optimizes the usage of spectrum over a long-time period, which we address in our future work. In this paper we focus on the possible gains in the achieved throughput when OPs have similar traffic patterns.

The core of the extensive LTE-A system level network simulator has been built according to the International Telecommunication Union's system level simulation guidelines \cite{WINII} and calibrated and rigorously evaluated in selected macro and microcell environments \cite{Pennanen}, \cite{Haataja}. The simulator is extended to incorporate indoor femtocells, calibrated and verified in \cite{befemto2} and previously utilized in \cite{Luoto1}, \cite{Luoto2}.

This paper is organized as follows. The system and link model is defined in Section \ref{sec:model}. Section \ref{sec:sharing} describes the channel quality information (CQI) model for CoPSS. In Section \ref{sec:algorithms}, the actual CoPSS algorithms are elaborated. Section \ref{sec:results} provides numerical results of the proposed CQI model and CoPSS algorithms. Finally, Section \ref{sec:conclusion} concludes the paper.

\section{System and Link Model} \label{sec:model}

Consider the downlink of an Orthogonal Frequency-Division Multiple Access (OFDMA) SBS network where $V$ SBSs are deployed. Each SBS has $N_\text{t}$ transmit antennas (Tx), which serve $U$ users each with $N_\text{r}$ receive antennas (Rx). The frequency domain resource consists of $N_\text{c}$ subcarriers, where 12 subcarriers are forming a PRB. It is assumed that the spectral allocations of the SBSs are orthogonal to the macro network layer and thus only the small cell traffic is modeled. Total system bandwidth is 10 MHz at 2 GHz center frequency and it is equally divided among the OPs.

In the system model SBSs form graphs. It is assumed that the SBSs communicate with each other if the distance is less than or equal to 50 meters.  Let $G_l = (V_l,E_l)$ denote the graph, where $l = [0,\dots,L]$ is the number of graphs, the number of vertices in the graph (in this case SBSs) is $V_l = \{v_0,\dots,v_n\}$ and the edges in the graph ($v_i$  is connected to SBS $v_j$) $E_{l} = \{(v_i,v_j)\}$. Let $\mathcal{K} = \{1,\dots,K\}$ denotes the set of OPs. We define function $\text{OP}(\cdot)$ which maps SBS $v_i$ to respective operator, i.e. $\forall v_i, \text{OP}(v_i) \in \mathcal{K}$.

If a graph has $n$ vertices we have an $n \times n$ matrix $\textbf{A}$ which is called an adjacency matrix. The matrix $\textbf{A}$ is defined by
\begin{equation}
\textbf{A}_{ij} = \begin{cases}
 \text{1} & \quad \text{if $\text{SBS}_i$ $\leftrightarrow$ $\text{SBS}_j$}\\
 \text{0} & \quad \text{otherwise}\\
\end{cases}
\end{equation}
where $i$ and $j$ are SBSs indices. When $\textbf{A}_{ij} =1$, $\text{SBS}_i$ and $\text{SBS}_j$ communicate successfully with one another. This matrix is formed by the central controller $v_0$ when each SBS reports its adjacent vector, which indicates the wireless connections of the SBS to other SBSs. OP/SBS is willing to share its bandwidth if they are not utilizing it fully, we define bandwidth utilization $\text{BWU}(v_i), \forall v_i \in (V_l-v_0)$, and each OP defines a sharing factor $S = [0,\dots,1]$ indicating how much they are willing to share if part of the bandwidth is free.

\begin{figure}
  \centering
    \includegraphics[trim = 0mm 0mm 0mm -4mm, clip, width=0.4\textwidth]{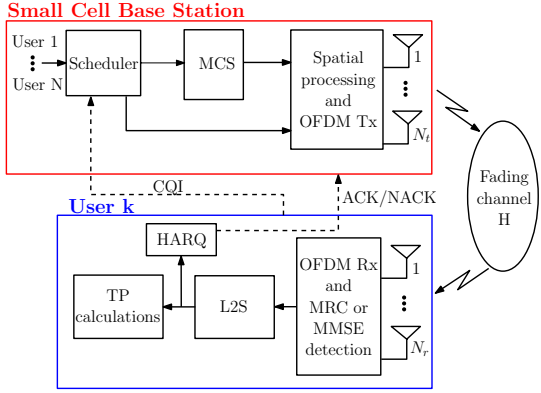}
    \caption{\label{link_model}Block diagram of the link model.}
\end{figure}

The link model between a SBS and a user is illustrated in Fig. \ref{link_model}. Because a link-to-system interface (L2S) is used in the simulations, coding/decoding and modulation/demodulation are omitted. Antenna gain, path loss and shadowing loss are calculated for all links. Each user is then paired to a SBS. A geometry-based stochastic channel model \cite{WINII,3GPP3} is used to model fast fading. Channel parameters are determined stochastically, based on the statistical distributions extracted from channel measurements \cite{itur2135}. SBS related assumptions for links are adopted from the \cite{befemto2}: all links are assumed to be non-line-of-sight (NLOS) and users are always inside buildings.

The link model starts from the scheduler that is responsible for resource allocation between users. Throughout simulations proportional fair scheduling is used. The scheduler utilizes CQI information transmitted by user. Based on the CQI information resource allocation is performed. The CQI provides information to the SBS about the link adaptation parameters. In the simulator, CQI is estimated from the received signal and for each user signal to interference and noise ratio (SINR) is calculated for every PRB. In order to model a practical closed loop system, periodic and delayed CQI is assumed. After scheduling, MCS selection is performed for scheduled users. The CQI modeling is explained in details in Section \ref{sec:sharing}. Finally, before the data is sent over the fading channel, transmitter side spatial and OFDM processing are performed. The cyclic prefix is assumed to be longer than the multipath delay spread, and thus inter-symbol-interference is not considered.

At the receiver, perfect frequency and time synchronization is assumed. Link-to-system mapping is performed using mutual information effective SINR mapping (MIESM) \cite{MIESM}. This significantly reduces the computational overhead compared with exact modeling of the radio links, while still providing sufficiently accurate results. In the link-to-system interface, SINR is calculated and it is mapped to corresponding average mutual information. Based on the MIESM value, the frame error probability (FEP) is approximated according to a predefined frame error rate (FER) curve of used MCS. Based on the FER, successful and erroneous frames can be detected, and hybrid automatic repeat request (HARQ) can take the control for retransmissions. Acknowledgement (ACK) or negative acknowledgement (NACK) message is sent back to the SBS to signal the success or failure of the transmission, respectively. When a predefined number of channel samples have been simulated the results are calculated.

\section{CQI modeling for Co-Primary Spectrum Sharing}\label{sec:sharing}
As mentioned in Section \ref{sec:model}, for each user the CQI is estimated from the received signal with SINR calculated for every PRB. When the SBS supports CoPSS, users have to calculate the CQI over the other operator's bandwidth. Here, user equipment (UE) is required to receive/request reference signals from the other operator's SBSs. In this case a user can only receive wideband reference signals from other OPs SBSs, because we assume that OPs are not willing to share operator specific reference signals. This means that users can only estimate if there are other OPs' SBSs nearby but they may not estimate the SINR accurately for each PRB when spectrum is shared. We propose a CQI model in which it is enough to know the BWU from other SBSs/OPs in order to make accurate CQI estimation.  

\begin{figure}
  \centering
    \includegraphics[trim = 0mm 0mm 0mm 0mm, clip, width=0.4\textwidth]{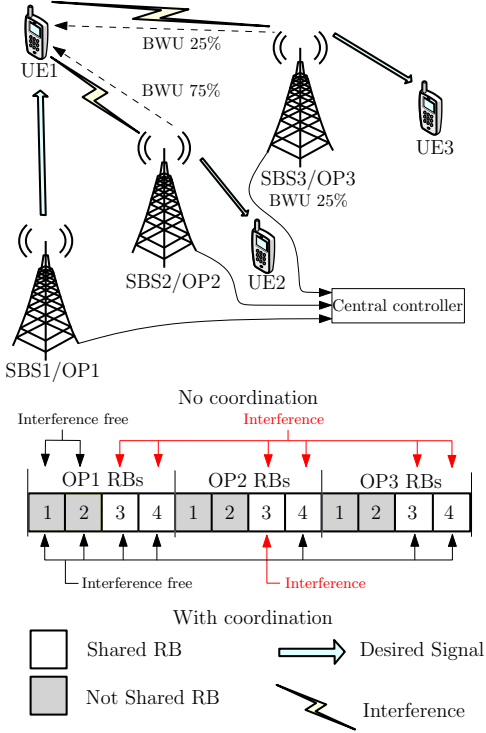}
    \caption{\label{CQI_modeling}CQI modeling with and without coordination between OPs.}
\end{figure}

Fig. \ref{CQI_modeling} shows an example of CQI modeling for UE1 when CoPSS is either supported or not (utilization of the central controller is explained in Section \ref{sec:algorithms}). Each operator has a bandwidth of 4 PRBs. UE1 is connected to SBS1/OP1, UE2 is connected to SBS2/OP2 and UE3 is connected to SBS3/OP3. Without any sharing UE1 is not aware of any interference in the network. Let us assume that 50\% of the bandwidth is shared, now UE1 can access OP2's and OP3's resources and vice versa. Without any coordination, when UE1 calculates the SINR for the CQI reporting it assumes that 50\% of its own OP's bandwidth is interference free and 50\% experiences interference from SBSs 2-3. UE1 also assumes that shared PRBs of other OPs are used when the SINR is calculated. The reason is that UE1 receives the wideband reference signals from other SBSs but it does not know whether the shared resources are used or not. This means that effectively a user makes a worst case estimate for the CQI, which gives the wort case performance that can be achieved if the resources are used as estimated. If users could make accurate estimation from other OPs bandwidth for each PRB this assumption is relaxed.

When there is coordination, each SBS receives the BWU of other SBSs. This information is included in the wideband reference signals that UEs are requesting from SBSs in the vicinity error free. In this example SBS2 and SBS3 transmit their BWU to UE1. Now UE1 can estimate the channel accurately and transmit an accurate CQI to SBS1. Without coordination UE1 would detect only two interference free PRBs, but with coordination seven interference free PRBs are detected.

In order for UE to predict which part of the bandwidth is not occupied, it has to know in which manner the SBS/OP allocates PRBs to its users. In order to minimize signaling overhead, in this work it is assumed that each SBS/OP starts allocating PRBs from the beginning or from the end of its bandwidth. Arbitrary allocations would require detailed resource allocation information exchange, significantly increasing the signaling overhead. When a UE knows how much bandwidth a SBS/OP is willing to share and what is the BWU, the UE can predict which part of the bandwidth is free and which part is occupied. This way the UE can estimate the CQI more accurately.

\section{Co-Primary Spectrum Sharing algorithms}\label{sec:algorithms}
With accurate CQI estimations, we propose three centralized and one decentralized algorithms for CoPSS. The proposed algorithms use moderate amount of shared information among OPs/SBSs and they do not require long iterative information exchange processes. Thus, the proposed CoPSS algorithms are practical.

In Algorithm \ref{random}, the free shared PRBs are randomly assigned to the SBSs in the building. The idea is that SBSs are connected to the central controller (as shown in Fig. \ref{CQI_modeling}) and there is no connection between SBSs resulting a single graph $G_l$ per building. This algorithm is time sensitive as there is a possibility that a randomly selected SBS from the graph can not exploit extra resources. Given that $v_{j'}$ is the selected SBS, the available free shared PRBs from OP $k$ to any SBS $v_j$ are given by:
\begin{equation}
{w}_{jk}=\begin{cases}\left\lfloor\min(W_k,S)\right\rfloor\times Q, & \text{if} ~ v_j=v_{j'}\\ 0 & \text{otherwise}, \label{allocation}
\end{cases}
\end{equation}
where $Q$ is the number of PRBs\footnote{Notation $\lfloor \cdot \rfloor$ defines the operation of round towards negative infinity.}, $S$ is the sharing factor and $W_k = 1-\max_{\substack{v_i \in \{v | \text{OP}(v)=k\}}}\big(\text{BWU}(v_i)\big)$ is the number of free PRBs at OP $k$. Thus, the total amount of free PRBs for SBS $v_j$ is $\sum_{k \in \mathcal{K}} w_{jk}$.
\begin{algorithm}
  \caption{Random sharing (centralized algorithm).}
        \begin{algorithmic}[1]
        \State Each SBS $v_i$ reports its BWU$(v_i)$ and sharing factor $S$ to the central controller $v_0$.
        \State $v_0$ picks $v_j \in (V_l-v_0)$ with probability $\frac{1}{|V_l-v_0|}$.
        \If {$\text{BWU}(v_j) = 1$}
            \State Allocate PRBs based on (\ref{allocation}).
        \Else
            \State $v_0$ does not allocate any resources.
        \EndIf
        \end{algorithmic}
        \label{random}
\end{algorithm}

In Algorithm \ref{equal}, the free PRBs are equally assigned to overloaded SBSs in the building. It is assumed that a SBS is overloaded if the whole bandwidth is utilized, i.e., BWU is one hundred percent. Sharing is performed in a centralized manner using the central controller. Therefore, we define new set $v^+ = \{v_i|\text{BWU}(v_i)=1\}$ which includes all the overloaded SBSs. Here, free shared PRBs from OP $k$ to SBS $v_j$ are
\begin{equation}
{w}_{jk}=\begin{cases}\left\lfloor\frac{1}{|v^+|}\min(W_k,S)\right\rfloor\times Q, & \text{if} ~ v_j=v_{j'}\\ 0 & \text{otherwise}, \label{allocation2}
\end{cases}
\end{equation}
and the total amount of free PRBs for SBS $v_j$ is $\sum_{k \in \mathcal{K}} w_{jk}$.

\begin{algorithm}
  \caption{Equal sharing (centralized algorithm).}
        \begin{algorithmic}[1]
        \State Each SBS $v_i$ reports its BWU$(v_i)$ and sharing factor $S$ to the central controller $v_0$.
        \State Central controller $v_0$ creates set $v^+$.
        \If {$\exists v^+ \neq \emptyset$}
            \State Allocate PRBs based on (\ref{allocation2}).
        \Else
            \State $v_0$ does not allocate any resources.
        \EndIf

        \end{algorithmic}
        \label{equal}
\end{algorithm}

Algorithm \ref{decentralized} aims to share resources equally between SBSs/OPs. The difference is that now SBSs are not connected to the central controller, but only to the SBSs in the vicinity, i.e sharing is done in a decentralized manner.  We let $\mathcal{N}(v_i)$ denotes the set of neighbor vertices of $v_i$ and from (\ref{neighbor}) we define two different sets, $\mathcal{N_\text{ol}}(v_i)$ for overloaded neighbors, and $\mathcal{N_\text{nol}}(v_i)$ for not overloaded neighbors,
\begin{eqnarray}
&\mathcal{N}(v_i)= \{v| v \in V_l, (v,v_i) \in E_l \} \label{neighbor}, \\
&\mathcal{N_\text{ol}}(v_i)= \{v| v \in \mathcal{N}(v_i), \text{BWU}(v)=1\} \label{ol},
\end{eqnarray}
and
\begin{eqnarray}
&\mathcal{N_\text{nol}}(v_i)= \{v| v \in \mathcal{N}(v_i), \text{BWU}(v)<1\}. \label{nol}
\end{eqnarray}

From (\ref{ol}), we define a set of OPs which are overloaded neighbors and rest of the OPs are not overloaded
\begin{equation}
\hat{\mathcal{K}}_i= \{\text{OP}(v)| \text{OP}(v) \in \mathcal{K}, v \in \mathcal{N_\text{ol}}\},
\end{equation}
and
\begin{equation}
\check{\mathcal{K}}_i= \mathcal{K}\backslash (\hat{\mathcal{K}_i} \cup \{\text{OP}(v_i)\}),
\end{equation}
respectively. Now we can define free shared PRBs $w_{j}$ from neighbors $\check{\mathcal{K}}_i$ to SBS $v_j$,
\begin{equation}
{w}_{j}=\sum_{\forall k \in \check{\mathcal{K}}_i} \left\lfloor \frac{\min\Bigg({1-\max_{\substack{v \in \mathcal{N}_\text{nol}(v_i)\\\text{OP}(v)=k}}\Big(\text{BWU}(v)\Big)},S\Bigg)\times Q}{|\hat{\mathcal{K}_{{i}}} \cup\{\text{OP}(v_i)\}|}\right\rfloor. \label{allocation3}
\end{equation}

\begin{algorithm}
  \caption{Connection based sharing (decentralized).}
        \begin{algorithmic}[1]
        \State Each $v_i$ reports BWU$(v_i)$ and sharing factor $S$ to all $v_j$ s.t $v_j \in V_l$ \text{and} $(v_i,v_j)\in E_l$.

        \State Each $v_i$ analyzes received reports.
        \If {BWU$(v_i)=1$}
            \If { $\mathcal{N}(v_i) = \emptyset $}

            \State $v_i$ allocates $|\mathcal{K}\backslash\text{OP}(v_i)|\times Q$ PRBs.

            \Else

            \State Allocate PRBs based on (\ref{allocation3}).

            \EndIf
        \EndIf
        \end{algorithmic}
        \label{decentralized}
\end{algorithm}

\begin{figure}
  \centering
    \includegraphics[trim = 0mm 0mm 0mm 0mm, clip, width=0.4\textwidth]{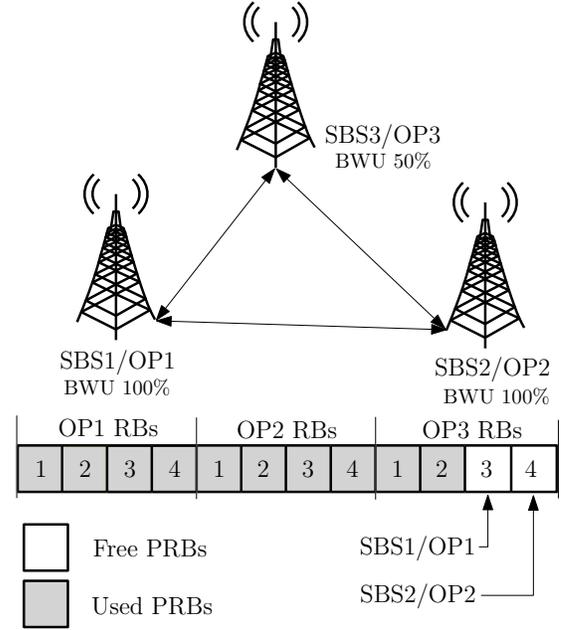}
    \caption{\label{algorithm3}Example sharing scenario for Algorithm \ref{decentralized}.}
\end{figure}

Fig. \ref{algorithm3} shows an example of how resources are shared. It is assumed that each OP has its own index and based on the index they know which part from the bandwidth resources can be taken from. In this example, 50\% of OP3 bandwidth is free. OP1 knows that half of the shared resources can be utilized in this case PRB3. Similarly OP2 knows that PRB4 can be utilized. Based on the number of overloaded SBSs/OPs, the unused portion of the bandwidth is divided equally. Allocations are interference free for each SBS/OP if the graph is a fully connected. For example a connection between SBS2 and SBS3 is not present, OP1 may utilize PRB3, but OP2 would see that OP3 is absent and may utilize 50\% of the resources, in this case PRBs 3 and 4. This means that PRB3 is utilized by OP1 and OP3 which results in interference.

In Algorithm \ref{graph}, aforementioned interference problem can be avoided as each SBS reports its connections to other SBS and BWU to the central controller. The central controller then forms an adjacent matrix. Utilizing the information from the adjacency matrix the central controller can generate interference free resource allocations as illustrated in the interference avoidance step in Algorithm \ref{graph}. Now we can define free shared PRBs $w_{j}$ from neighbors to SBS $v_j$ as

\begin{equation}
{w}_{j}=\sum_{\forall k \in \check{\mathcal{K}}_i} \left\lfloor \frac{\min\Bigg({1-\max_{\substack{v \in \mathcal{N}_\text{nol}(v_i)\\\text{OP}(v)=k}}\Big(\text{BWU}(v)\Big)},S\Bigg)\times Q}{|\mathcal{N_\text{ol}}(v_i)|+1}\right\rfloor. \label{allocation4}
\end{equation}

\begin{algorithm}
  \caption{Connection based sharing (centralized).}
        \begin{algorithmic}[1]
        \State Each SBS reports the bandwidth utilization (BWU), sharing percentage and adjacent vector to central controller.

        \State Central controller forms adjacent matrix \textbf{A} and adjacent matrix of overloaded SBSs \textbf{\^{A}}.

            \If {$\textbf{\^{A}} \in \mathbb{R}$}
            \State $\exists~ v_i$ s.t. BWU$(v_i)=1$.

                \If { $\mathcal{N}(v_i) = \emptyset $}

                \State $v_i$ allocates $|\mathcal{K}\backslash\text{OP}(v_i)|\times Q$ PRBs.

                \Else

                \State Allocate PRBs based on (\ref{allocation3}).

                \EndIf
            \Else~{$\hat{n}$ overloaded SBSs, \textbf{\^{A}} $\in \mathbb{R}^{\hat{n}\times\hat{n}}$}
            \State \emph{PRB allocation step}
            \For {$~v_i = v_1:v_{\hat{n}}$}

             \State Allocate PRBs $w_i$ based on (\ref{allocation4}).
             \State Remove allocated PRBs  from the available free
             \Statex \hspace{6ex} resources.

            \EndFor
            \State \emph{Interference avoidance step}

                \For {$~w_i = w_1:w_{\hat{n}}$}

                \State $w_i \leftarrow w_i \backslash (w_i \cap w_j), j=1,\ldots,\hat{n}, j\neq i$.

            \EndFor

            \EndIf
        \end{algorithmic}
        \label{graph}
\end{algorithm}

The round trip-delay in the coordination methods is 5 ms and it is assumed that each OP uses the same maximum allowed sharing percentage\footnote{Percentages could be different but results are easier to analyze when same sharing percentage is used because we do not have to look at the gains achieved for each OP individually.}. Backhaul links between SBSs or connections to the central controller are assumed to be ideal.

\section{System Level Performance Results}\label{sec:results}
System level simulations are particularly useful for studying network related issues such as resource allocation, interference management and mobility management. In this work, a multi-operator LTE-A system level simulator is used to model a cellular network consisting of an indoor SBS with multiple OPs.

\begin{figure}
  \centering
    \includegraphics[trim = 0mm 0mm 0mm 0mm, clip, width=0.5\textwidth]{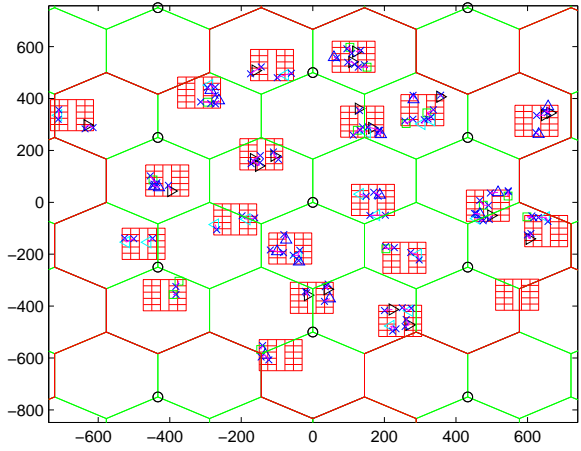}
    \caption{\label{Layout}The network layout of the simulator.}
\end{figure}

\begin{figure}
  \centering
    \includegraphics[trim = 0mm 0mm 0mm 10mm, clip, width=0.5\textwidth]{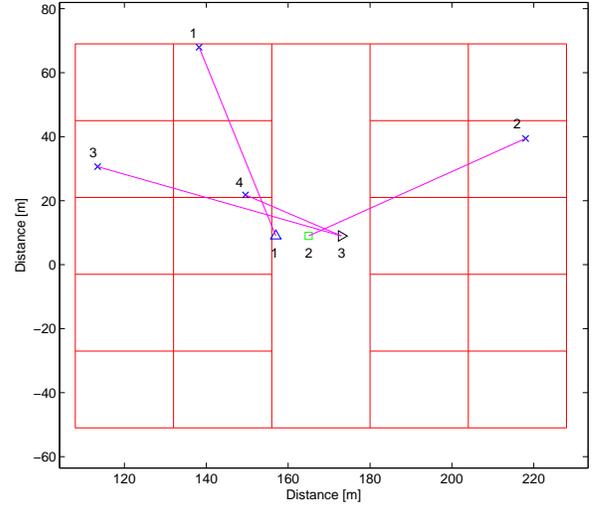}
    \caption{\label{Even_layout}Small cell layout where base stations are colocated in fixed central positions.}
\end{figure}

\begin{figure}
  \centering
    \includegraphics[trim = 10mm 5mm 10mm 10mm, clip, width=0.5\textwidth]{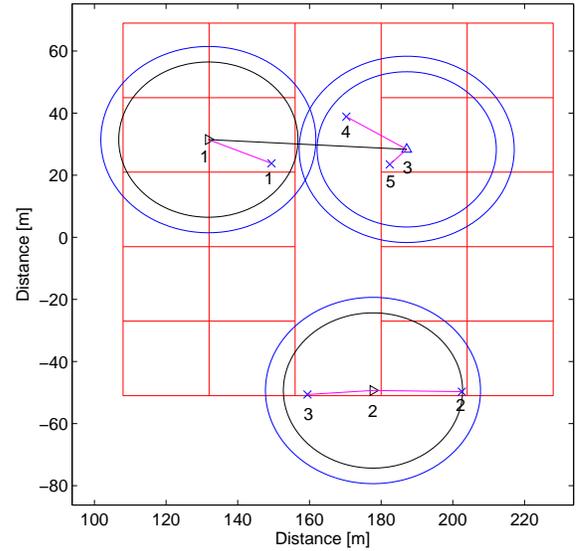}
    \caption{\label{Random_layout2}Small cell layout where base stations are located randomly.}
\end{figure}

The simulator uses a hexagonal macro layout which includes 21 sectors and in each sector there is a building of the size 120 m x 120 m as shown in Fig. \ref{Layout}. Small cell layouts are shown in Figs. \ref{Even_layout} (fixed layout) and \ref{Random_layout2} (random layout). The building has one open corridor across it and in total 20 rooms, size 24 m x 24 m. Internal wall attenuation is 5 dB per wall.

When the locations of the SBSs are fixed, they are placed in the center of the building and each OP has one SBS per building, users are evenly distributed and each of them is connected to the own OP's SBS. When SBSs are randomly distributed, the number of SBSs in the building is based on deployment probability. In this layout users are located a maximum of 20 m from the SBS (inner circle). In both layouts, the number of users connected to each SBS varies between one and two. The main reason to use two different layouts is to illustrate the applicability of the CoPSS concept in both planned (fixed) and unplanned (random) SBS deployment scenarios.

In the simulations, two different traffic models are used, full buffer and continuous constant rate transmission. With continuous constant rate transmission, two different target bit-rates are used; 4 Mb/s target rate is referred to as multimedia stream (e.g., on-demand video service) and 1 Mb/s target rate is referred to as constant rate (e.g., users with a limited speed data connection).

Table \ref{params} summarizes some simulation parameters and assumptions which are used through simulations. Traffic in the network is constant, and movement of users is not modeled. This means that delay does not have a big impact on the performance because SBS resource allocation stays quite consistent throughout simulations.

\begin{table}
\centering
\caption{Simulator parameters and assumptions.}%
\label{params}
\begin{tabular}{|p{3cm}|p{4cm}|}
  \hline
  \textbf{Parameter} & \textbf{Assumption} \\ \hline
      Duplex mode & FDD \\ \hline
      System bandwidth & 10 MHz (divided equally between OPs) \\ \hline
      Number of PRBs & 16 per SBS/OP \\ \hline
      Number of users & 1-2 user per SBS \\ \hline
      Antenna configurations & 1 Tx, 2 Rx\\ \hline
      Receivers & MRC \\ \hline
      HARQ & Chase combining \\ \hline
      SBS transmission power & 20 dBm \\ \hline
      Feedback CQI period & 6 ms \\ \hline
      Feedback CQI delay & 2 ms \\ \hline
      Traffic models & \textbf{Full buffer} (10\% full buffer traffic) and \textbf{Continuous constant rate transmission} (50\% constant rate traffic and 40\% multimedia stream traffic)\\ \hline
      Internal wall attenuation & 5 dB \\
  \hline
\end{tabular}
\end{table}

Algorithms \ref{random}-\ref{decentralized} are used for both network layouts and Algorithm \ref{graph} is only used for the random layout. For each CoPSS algorithm, the target is to allocate only shared resources that are unused within the network. It should be noted that CoPSS is highly sensitive to the network load and to different traffic types. In these simulations the network load (1-2 users per SBS) is relatively low, however for a high network load (i.e., all resources utilized) Algorithms \ref{random} - \ref{decentralized} in the fixed layout and Algorithms \ref{random} - \ref{equal} in the random layout would not provide any gain because only unused PRBs are shared between OPs/SBSs. The reason for zero gains in the fixed layout is that when all resources are utilized, and all OPs are colocated (\ref{allocation}), (\ref{allocation2}) and (\ref{allocation3}) are always equal to zero. Similarly, (\ref{allocation}) and (\ref{allocation2}) are equal to zero in the random layout. However, the graphs in the random layout can have OPs that are not colocated and thus, (\ref{allocation3}) and (\ref{allocation4}) provide non-zero gains.

\subsection{CQI coordination for CoPSS}
\begin{figure}
  \centering
    \includegraphics[trim = 0mm 0mm 0mm 0mm, clip, width=0.5\textwidth]{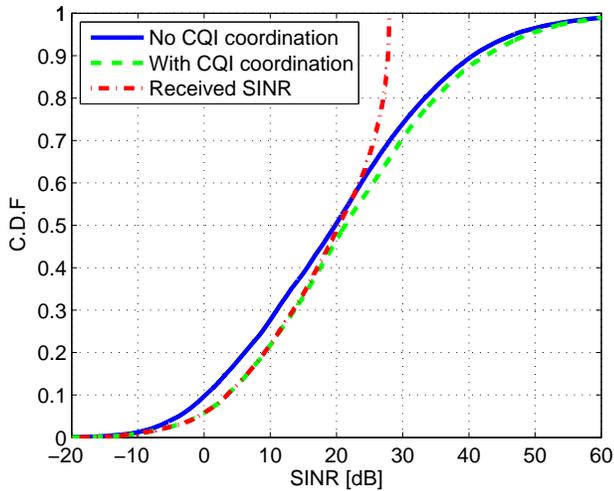}
    \caption{\label{Equal_CQI_analysis_SINR}UEs SINR with and without CQI coordination and in the receiver when the fixed layout is used.}
\end{figure}

Fig. \ref{Equal_CQI_analysis_SINR} shows cumulative distribution functions (CDFs) of SINR as estimated for the CQI with and without coordination, and as experienced at the receiver, when 50\% bandwidth is shared. The SINR in the CQI and in the receiver is the mean SINR over the allocated PRBs. It is assumed that UE can report the PRB based CQI information and SBS then averages out the SINR with allocated PRBs and then selects one MCS level that is used for the transmission. The CDF shows that when the users are able to receive information about the bandwidth utilization of other SBSs/OPs in the vicinity SINR increases 3 dB (between 0 dB and 15 dB).

In Fig. \ref{Equal_CQI_analysis_SINR}, it can be seen that the SINR at the receiver saturates around 28 dB. The reason is that in the receiver side error vector magnitude (EVM) is used to model hardware imperfection, which is assumed to have a value of 4\%. The EVM error to the received SINR can be written as:
\begin{equation}
SINR_\text{out} = 1/\Big((1/SINR_\text{in})+(EVM_\text{\%}/100)^2\Big)
\end{equation}
where $SINR_\text{in}$ is the received SINR in linear scale and $EVM_\text{\%}$ is the percentage EVM.

\begin{figure}
  \centering
    \includegraphics[trim = 0mm 0mm 0mm 0mm, clip, width=0.5\textwidth]{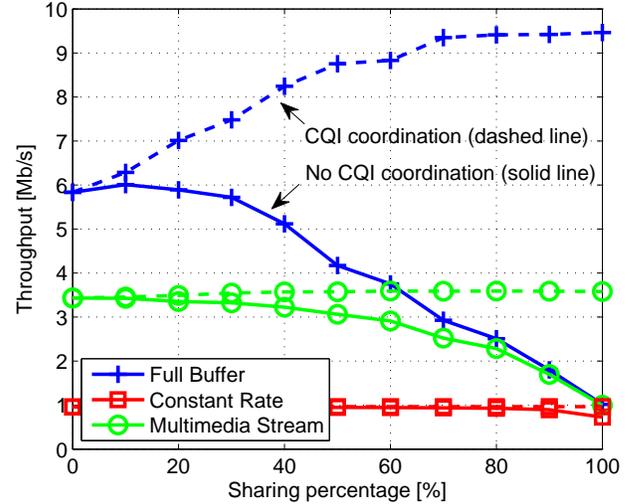}
    \caption{\label{Equal_CQI_analysis_SP}Throughput with sharing percentage, with and without CQI coordination when small cells are in fixed positions.}
\end{figure}

Fig. \ref{Equal_CQI_analysis_SP} shows the mean throughput when the sharing percentage increased from 0\% to 100\% with and without CQI coordination. The used CoPSS algorithm is equal sharing. The SBSs are fixed in the center of the building. Results show that when the sharing percentage is increased and there is no CQI coordination, achieved mean throughput starts to decrease particulary for the full buffer and multimedia stream users. When 100\% of the bandwidth is shared full buffer and multimedia stream users achieve throughput of 1 Mb/s, i.e. 5 Mb/s loss for the full buffer users and 2.5 Mb/s loss for the multimedia users compared with the case when 0\% of the bandwidth is shared. When 100\% of the bandwidth is shared the equal CoPSS provides a 3.8 Mb/s increase in the mean throughput for full buffer users when compared to the case without the CoPSS. When the sharing percentage is increased and there is CQI coordination, the multimedia stream and the constant rate users do not achieve any gain in mean throughput because most of the users can achieve the target bit rate, i.e. achieved throughput gain averages out. The CDFs of throughput are analyzed in Section \ref{sec:results}.

\begin{figure}
  \centering
    \includegraphics[trim = 0mm 0mm 0mm 0mm, clip, width=0.5\textwidth]{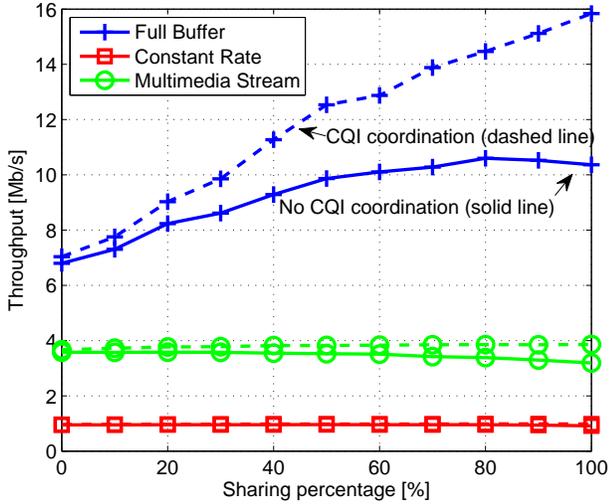}
    \caption{\label{Random_CQI_analysis_SP}Throughput with sharing percentage, with and without CQI coordination when small cells are randomly distributed.}
\end{figure}

Fig. \ref{Random_CQI_analysis_SP} shows the mean throughput results when SBSs are randomly distributed in the building. Results show that when the sharing percentage is increased and there is no CQI coordination, the achieved mean throughput for full buffer users starts to increase, but for the multimedia stream users there is reduction in throughput. The reason is that for full buffer users the huge increase in available PRBs outweighs the loss by underestimated CQI, but for the lower data rate users the underestimation of the CQI leads to throughput reduction when the sharing percentage is increased. When 100\% of the bandwidth is shared with CQI coordination the equal CoPSS provides a 9.0 Mb/s increase in the mean throughput for full buffer users, when compared with the case without sharing.

These results show that coordination is needed between OPs if CoPSS is supported in the network. Without coordination, quality of the service can not be guaranteed and CoPSS can even result in a loss in performance. In the rest of the discussion it is assumed that there is a coordination between the OPs/SBSs in the vicinity of one another.

\subsection{CoPSS in the fixed layout}
First the different CoPSS algorithms are analyzed in the fixed network given in Fig. \ref{Even_layout} where all the SBS are collocated and interconnected, i.e there is a simple complete graph $G_l$ per building. In this network layout, it is crucial that simultaneous use of the shared PRBs is avoided. Because the SBSs are close to each other the serving signal and the interference signal would have approximately the same strength, leading to a high FER. Decentralized sharing and equal sharing in the fixed layout should provide similar performance (all the SBS are collocated and interconnected) if there is a common protocol between OPs defining how shared resources can be utilized.

\begin{figure}
  \centering
    \includegraphics[trim = 0mm 0mm 0mm 0mm, clip, width=0.5\textwidth]{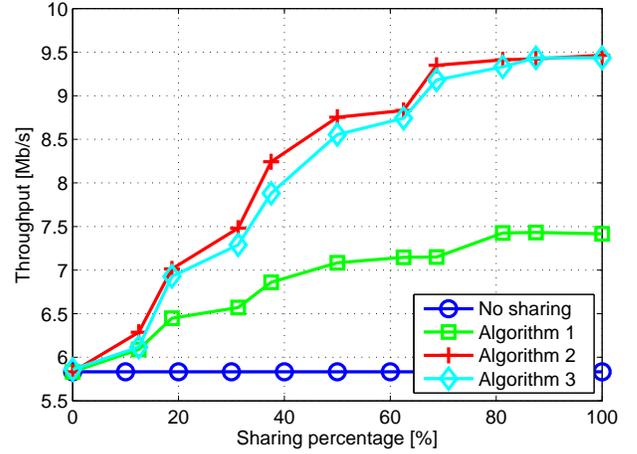}
    \caption{\label{Equal_algorithm_analysis_SP}Comparison of the different CoPSS algorithms in the fixed network layout.}
\end{figure}

Fig. \ref{Equal_algorithm_analysis_SP} shows the mean throughput of full buffer users for each CoPSS algorithms with sharing percentage from 0\% to 100\%. When 0\% of the bandwidth is shared mean throughput is 6.0 Mb/s. It can be clearly seen that all the CoPSS algorithms result in throughput gains, increasing with the sharing percentage. As expected, Algorithm 1 provides the lowest gain, a 1.5 Mb/s increment to mean throughput when 100\% of the bandwidth is shared. Algorithm 2 provides a 3.7 Mb/s gain. As mentioned in Section \ref{sec:algorithms}, Algorithm 2 and Algorithm 3 provide very similar performance because each SBS has the same knowledge as the central controller.

\begin{figure}
  \centering
    \includegraphics[trim = 0mm 0mm 0mm 0mm, clip, width=0.5\textwidth]{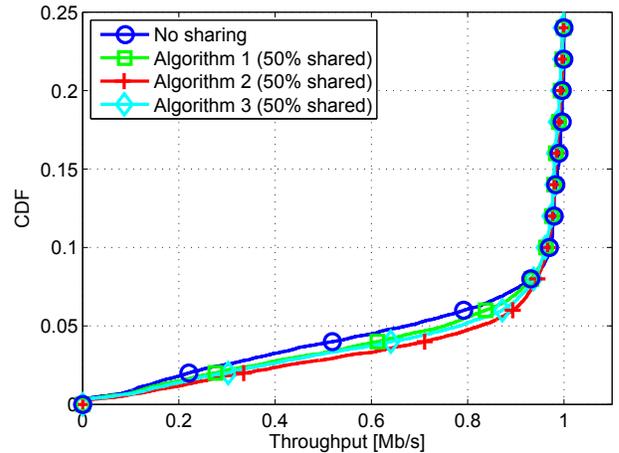}
    \caption{\label{Equal_algorithm_analysis_throughput1}CDF of Constant Rate user throughput for different CoPSS algorithms in the fixed network layout.}
\end{figure}

\begin{figure}
  \centering
    \includegraphics[trim = 0mm 0mm 0mm 0mm, clip, width=0.5\textwidth]{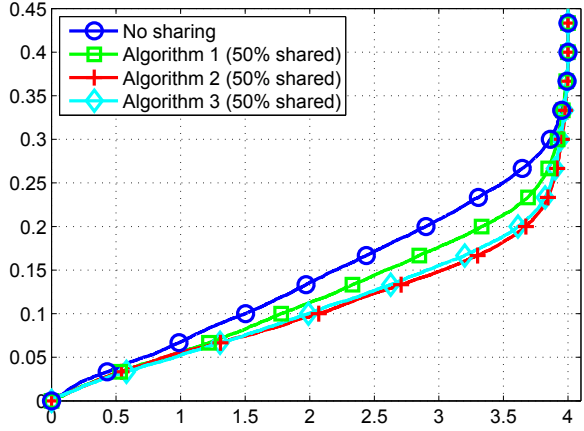}
    \caption{\label{Equal_algorithm_analysis_throughput2}CDF of Multimedia Stream user throughput for different CoPSS algorithms in the fixed network layout.}
\end{figure}

CoPSS provides substantial gains for full buffer users. Figs. \ref{Equal_CQI_analysis_SP} and \ref{Random_CQI_analysis_SP} imply that only full buffer users achieve some gain from CoPSS. Figs. \ref{Equal_algorithm_analysis_throughput1} and \ref{Equal_algorithm_analysis_throughput2} show the CDF of throughput for the constant rate and multimedia stream users when 50\% of the bandwidth is shared. From these figures it can be seen that all the CoPSS methods provide gain over the case when the spectrum is not shared, for users with all traffic types. Fig. \ref{Equal_algorithm_analysis_throughput2} shows for example at the 20\% point on the CDF there is a 0.7 Mb/s gain in throughput when Algorithm 3 is compared to no spectrum sharing.

\begin{figure}
  \centering
    \includegraphics[trim = 0mm 0mm 0mm 0mm, clip, width=0.5\textwidth]{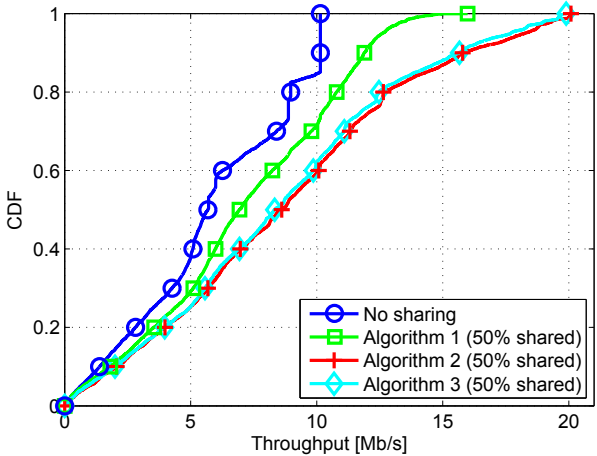}
    \caption{\label{Equal_algorithm_analysis_throughput3}CDF of Full Buffer user throughput for different CoPSS algorithms in the fixed network layout.}
\end{figure}

Fig. \ref{Equal_algorithm_analysis_throughput3} shows the CDF of throughput for full buffer users. The theoretical maximum throughput of a user (when the SBS is only serving one user and the highest MCS is used) is around 10 Mb/s. The CDF shows that a user will achieve a throughput of 10 Mb/s with a probability of 18\%. For example at the 90\% point on the CDF achieved gains are; 2.0 Mb/s for Algorithm 1, 5.8 Mb/s for Algorithm 2, and 5.7 Mb/s for Algorithm 3. When compared to the theoretical maximum throughput of 10 Mb/s without sharing, the gain is significant. Table \ref{gains1} summarizes the achievable gains of cell edge users (5\% from CDFs) when the CoPSS is supported.

\begin{table}
\centering
\caption{Cell edge user throughput gain [Mb/s] with the CoPSS in the fixed layout.}%
\label{gains1}
\begin{tabular}{|c|c|c|c|}
  \hline
  \textbf{Algorithm} & \textbf{Constant rate} & \textbf{Multimedia stream} & \textbf{Full buffer}  \\ \hline
      1 & 0.080  & 0.180 & 0.096 \\ \hline
      2 & 0.154  & 0.159 & 0.327 \\ \hline
      3 & 0.110  & 0.265 & 0.261 \\ \hline

  \hline
\end{tabular}
\end{table}

\subsection{CoPSS in the random layout}
In the random layout (Fig. \ref{Random_layout2}), a connection is formed when the coverage area of two SBSs overlap with one another, in this case a maximum distance 20 m + 5 m range is used. It is assumed that within this distance if same resources are used, users will experience high interference from neighboring SBSs. When users are within 20 m range, the SBS is working as a local hotspot, and allows for higher data rates and spectral effciency resulting in better user experience. When a SBS does not detect the presence of any SBS belong to a particular OP within its detection range, it assumes those OPs' resources to be free and exploitable.

\begin{figure}
  \centering
    \includegraphics[trim = 0mm 0mm 0mm 0mm, clip, width=0.5\textwidth]{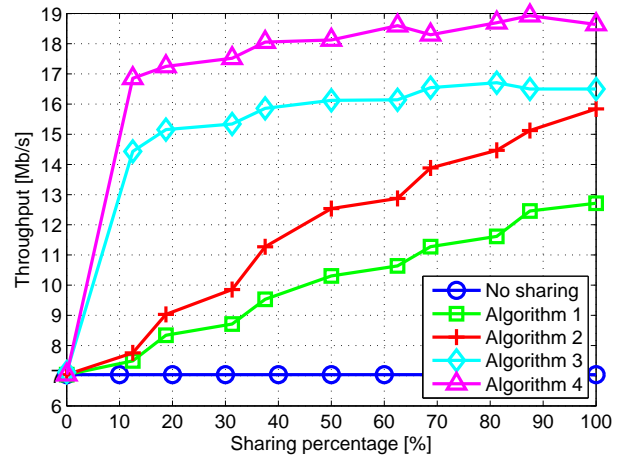}
    \caption{\label{Random_algorithm_analysis_SP}Comparison of the different CoPSS algorithms in the random network layout.}
\end{figure}

Fig. \ref{Random_algorithm_analysis_SP} shows the mean throughput of full buffer users for each CoPSS algorithm with sharing percentage from 0\% to 100\%. When the results are compared with the fixed layout results it can be clearly seen that the achieved rates are higher because users are now closer to SBS. When 0\% of the bandwidth is shared mean throughput is 7.0 Mb/s. Algorithm 1 provides the lowest gain, a 5.7 Mb/s improvement to mean throughput when 100\% of the bandwidth is shared, while Algorithm 2 provides a 8.8 Mb/s gain. The achieved gain from Algorithm 3 is 9.6 Mb/s, and the Algorithm 4 results in the highest gain 11.6 Mb/s. The reason of Algorithm 4 providing higher gains compared to Algorithm 3 is explained in Section \ref{sec:algorithms} and Fig. \ref{algorithm3}.

\begin{figure}
  \centering
    \includegraphics[trim = 0mm 0mm 0mm 0mm, clip, width=0.5\textwidth]{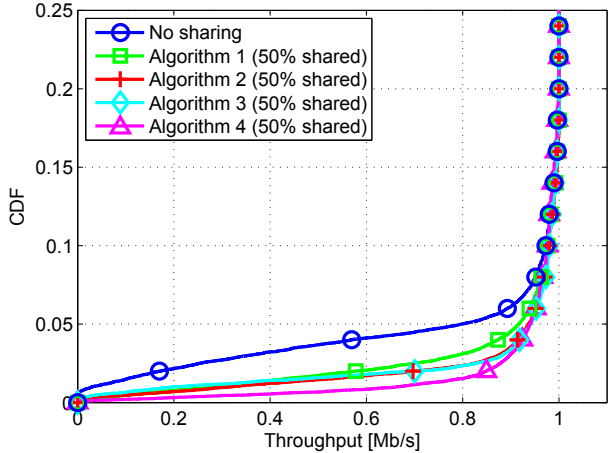}
    \caption{\label{Random_algorithm_analysis_throughput1}CDF of Constant Rate user throughput for different CoPSS algorithms in the random network layout.}
\end{figure}

\begin{figure}
  \centering
    \includegraphics[trim = 0mm 0mm 0mm 0mm, clip, width=0.5\textwidth]{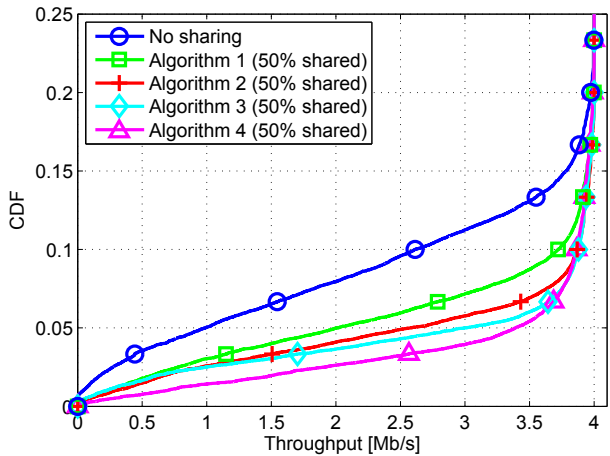}
    \caption{\label{Random_algorithm_analysis_throughput2}CDF of Multimedia Stream user throughput for different CoPSS algorithms in the random network layout.}
\end{figure}

In the fixed layout, although there are no significant gains for low-data rate users, all the CoPSS methods result higher gains over the scenario with no spectrum sharing. In the random layout the gains from CoPSS are higher than in the fixed layout. Fig. \ref{Random_algorithm_analysis_throughput1} and \ref{Random_algorithm_analysis_throughput2} show the CDF of throughput for the constant rate and multimedia stream users when 50\% of the bandwidth is shared. In Fig. \ref{Random_algorithm_analysis_throughput2}, there is 10\% probability of achieving less than 2.6 Mb/s, which is reduced to 4\% in the case of sharing (Algorithm 4).

\begin{figure}
  \centering
    \includegraphics[trim = 0mm 0mm 0mm 0mm, clip, width=0.5\textwidth]{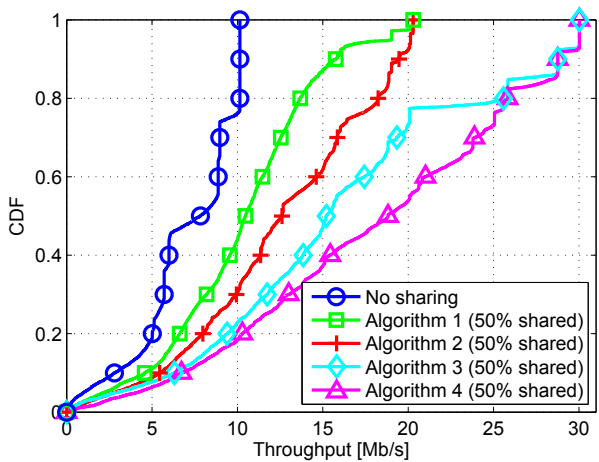}
    \caption{\label{Random_algorithm_analysis_throughput3}CDF of Full Buffer user throughput for different CoPSS algorithms in the random network layout.}
\end{figure}

Fig. \ref{Random_algorithm_analysis_throughput3} shows the CDF of throughput for the full buffer users. At the 50\% point on the CDF the achieved rates are: 7.8 Mb/s for No sharing, 10.5 Mb/s for Algorithm 1, 11.6 Mb/s for Algorithm 2, 15.2 Mb/s for Algorithm 3 and Algorithm 4 18.8 Mb/s. The achieved gains are significant compared to the no spectrum sharing scenario with the theocratical maximum of 10 Mb/s. Table \ref{gains2} summarizes the gains of cell edge users when the CoPSS is supported.

\begin{table}
\centering
\caption{Cell edge user throughput gain [Mb/s] with the CoPSS in the random layout.}%
\label{gains2}
\begin{tabular}{|c|c|c|c|}
  \hline
  \textbf{Algorithm} & \textbf{Constant rate} & \textbf{Multimedia stream} & \textbf{Full buffer}  \\ \hline
      1 & 0.117  & 1.022 & 1.171 \\ \hline
      2 & 0.137  & 1.594 & 1.811 \\ \hline
      3 & 0.144  & 2.008 & 1.869 \\ \hline
      4 & 0.141  & 2.403 & 3.129 \\
  \hline
\end{tabular}
\end{table}

\subsection{CoPSS behavior with higher network load}
As discussed earlier, a higher network load limits the achievable throughput gains using CoPSS. Table \ref{network_load} shows the mean achieved throughput of full buffer users, for an increasing network load, when 100\% of each OPs bandwidth is shared. The results clearly show that the gain in average throughput when utilizing CoPSS decreases with the network load. However, utilizing CoPSS does result in non-negligible increased throughput even in the case of a high network load.
\begin{table}
\centering
\caption{Achieved mean throughput [Mb/s] with CoPSS in the random layout with different network loads.}%
\label{network_load}
\begin{tabular}{|c|c|c|c|c|c|}
  \hline
  \textbf{Users} & \textbf{No sharing} & \textbf{Algo. 1} & \textbf{Algo. 2} & \textbf{Algo. 3} & \textbf{Algo. 4} \\ \hline
      1-2 & 7.04  & 12.71 & 15.84 & 16.50 & 18.64 \\ \hline
      1-4 & 4.84  & 9.11  & 10.68 & 11.64 & 13.07     \\ \hline
      1-6 & 3.34  & 6.27  & 7.30  & 9.08 & 9.69    \\
  \hline
\end{tabular}
\end{table}

Although Algorithm \ref{random} is totally random, it exhibits significant throughput gains compared to the no sharing method. Thus, even a simple CoPSS can help to improve capacity in SBS network scenarios. This type of sharing does not guarantee that resources are shared equally between SBSs/OPs during one time instant, but each SBS has an equal chance to be chosen. Algorithms \ref{equal} and \ref{decentralized} in the fixed network layout provide similar performance because all the SBS are collocated and interconnected. Generally, if all SBSs are connected, our proposed algorithm provides substantial throughput gain without central controller.

In the random layout, Algorithms \ref{equal} and \ref{decentralized} exhibit different performance. In this case, the decentralized Algorithm \ref{decentralized} outperforms the centralized Algorithm \ref{equal}. This is due to the reason that the decentralized algorithm does not share resources equally within each building, but resources are shared between SBSs that are within communication range of one another. In this case, an isolated SBS achieves significant gains in throughput even for a low sharing percentage. When Algorithm \ref{decentralized} and Algorithm \ref{graph} are compared, the centralized algorithm provides better performance as explained in Section \ref{sec:algorithms}. The decentralized Algorithm \ref{decentralized} provides substantial gains as compared to the no sharing case, with an average gain of more than 120\%. However, the centralized Algorithm \ref{graph} only results in additional average gain of 12\% over Algorithm \ref{decentralized}. Given that the performance of the decentralized Algorithm \ref{decentralized} is so close to that of the centralized Algorithm \ref{graph}, we come to the conclusion that Algorithm \ref{decentralized} is the most suitable for all the aforementioned scenarios.

The proposed algorithms do not require complex computation, or extensive signaling between SBSs. Algorithms \ref{equal} - \ref{graph} reach stable point quickly, and the only delay is the coordination delay between SBSs/OPs. This is because there is no requirement for iterative information exchange between SBSs/OPs, due to the common rules between SBSs/OPs, which determine how spectrum is shared.

\section{Conclusion}\label{sec:conclusion}
We have proposed and evaluated four different approaches toward co-primary multi-operator spectrum sharing in small cell indoor environment with mixed traffic distribution. The framework has been established under the LTE-A compliant system simulation platform where the system throughput performance has been rigorously assessed. Provided numerical results confirm the high potential co-primary spectrum sharing can offer to increase system throughput in the multi-operator setting. The results reveal the utmost importance of channel quality signaling among OPs in order to take full advantage of shared resources. It was also shown that, the connection based centralized and decentralized algorithms outperform simpler random and equal sharing schemes. This paper is a foundation for further studies. In our future work, we will study CoPSS with time variant network traffic, develop algorithms that ensure long term fairness between OPs and we will consider the economic part of the spectrum sharing in more detail.

\ifCLASSOPTIONcaptionsoff
  \newpage
\fi

\begin{small}
\bibliographystyle{IEEEtran}
\bibliography{IEEEabrv,Referencelist}
\end{small}

\end{document}